\documentclass[manuscript]{aastex}
\usepackage{amsmath}

\shorttitle{Tachyon-Dominated Cosmology and Type Ia Supernovae}
\shortauthors{Kramer and Redmount}

\begin{document}
\vskip -1.0in
\title{Testing Tachyon-Dominated Cosmology with Type Ia Supernovae}

\author{Samuel H.~Kramer}
\affil{University of Wisconsin--Madison\\
	Department of Physics\\
	1150 University Avenue\\
	Madison, Wisconsin~~53706--1390~~USA}
\email{shkramer@wisc.edu}


\author{Ian H.~Redmount}
\affil{Saint Louis University\\
	Department of Physics\\
	3511 Laclede Avenue\\
	Saint Louis, Missouri~~63103--2010~~USA}
\email{ian.redmount@slu.edu}

\begin{abstract}
An open or hyperbolic Friedmann-Robertson-Walker spacetime dominated by
tachyonic dark matter can exhibit an ``inflected'' expansion---initially
decelerating, later accelerating---similar but not identical to that of
now-standard $\Lambda$CDM models dominated by dark energy.  The features
of the tachyonic model can be extracted by fitting the redshift-distance
relation of the model to data obtained by treating Type~Ia supernovae as
standard candles. Here such a model is fitted to samples of~186 and~1048
Type~Ia supernovae from the literature.  The fits yield values of
$H_0=(66.6\pm1.5)~\hbox{km/s/Mpc}$ and $H_0=(69.6\pm0.4)~\hbox{km/s/Mpc}$,
respectively, for the current-time Hubble parameter, and
$t_0=(8.35\pm0.68)~\hbox{Gyr}$ and $t_0=(8.15\pm0.36)~\hbox{Gyr}$,
respectively, for the comoving-time age of the Universe.  Tests of the
model against other observations will be undertaken in subsequent works.
\end{abstract}


\section{Introduction}\label{sec01}

Adhering strictly to Einsteinian mechanics, it is possible to construct
a consistent mechanics for tachyons---faster-that-light particles---and
kinetic theory for a tachyon gas.~\citep{star2022}  The resulting equation
of state implies that a Friedmann-Robertson-Walker spacetime with mass/energy
content dominated by such a gas can exhibit a variety of evolutionary
behaviors.  In particular, for a suitable choice of parameters, an open
or hyperbolic tachyon-dominated model undergoes ``inflected''
expansion---decelerating from an initial singularity to a minimum expansion
rate, subsequently accelerating---such as is suggested by a variety of recent
observational evidence.

The utility of such a model is of course to be determined by comparing its
predictions with observational data.  One observational pillar of modern
cosmology is the distance/redshift data for Type~Ia supernovae.  These
objects can serve as ``standard candles'':  Their luminosities (absolute
magnitudes) can be determined from their light curves; hence, their
luminosity distances can be determined from their apparent magnitudes.
The distance-redshift relation for the open tachyon-dominated model can be
expressed analytically in closed form.  Fitting this relation to suitable
sets of data via $\chi^2$-minimization determines the best-fit parameters
of the model and all its features.

In this work samples of~186 and~1048 redshift/distance data points are curated
from the literature.  The model parameters and their uncertainites are
determined via a Monte Carlo calculation:  Multiple iterations of a
Levenberg-Marquardt $\chi^2$~minizmation are carried out, with normally
distributed variations introduced into the data.  The predictions of the
model and the goodness-of-fit to the data are thus evaluated.

The open tachyon-dominated cosmological model is detailed in Sec.~\ref{sec02}.
The data sets and their sources are described in Sec.~\ref{sec03}.  The fitting
of the model to the data and the results obtained are shown numerically and
graphically in Sec.~\ref{sec04}.  Conclusions and directions for future work
are described in Sec.~\ref{sec05}.

\vfil\eject
\section{Open Tachyon-Dominated Cosmology}\label{sec02}

The open tachyon-dominated cosmological model of interest here is described
by line element
\begin{subequations}
\begin{equation}
\label{eq01a}
ds^2=-c^2\,dt^2+a^2(t)\left[d\chi^2+\sinh^2\chi\,\left(d\theta^2
+\sin^2\theta\,d\phi^2\right)\right]\ ,
\end{equation}
with comoving time coordinate~$t\in[0,+\infty)$, angular coordinates
$\chi\in[0,+\infty)$, $\theta\in[0,\pi]$, and $\phi\in[0,2\pi)$, and
scale factor or curvature radius~$a(t)$ given parametrically by
\begin{equation}
\label{eq01b}
a(\eta)=\frac{A}{c}\,\sinh\eta-\frac{B}{c^2}\,(\cosh\eta-1)
\end{equation}
and
\begin{equation}
\label{eq01c}
t(\eta)=\frac{A}{c^2}\,(\cosh\eta-1)-\frac{B}{c^3}\,(\sinh\eta-\eta)\ ,
\end{equation}
\end{subequations}
in terms of \textit{conformal time} parameter $\eta\in[0,+\infty)$.
This spacetime geometry is driven, per the Einstein Field Equations,
by energy density
\begin{subequations}
\begin{align}
\rho(a)&=\frac{3c^2}{8\pi G}\left(\frac{A^2}{a^4}-\frac{2B}{a^3}\right)
\label{eq02a}\\
&=\frac{(\rho_0+{\cal M}_0c^2)a_0^4}{a^4}
-\frac{{\cal M}_0c^2a_0^3}{a^3}\ ,\label{eq02b}
\end{align}
\end{subequations}
with $G$ the Newton constant, and $\rho_0$ the energy density and ${\cal M}_0$
an invariant-mass density at fiducial time~$t_0$ (e.g., the present time),
at which the scale factor takes value~$a_0$.  This density follows from the
equation of state appropriate to a thermal ensemble of free, noninteracting
(i.e., dark-matter) tachyons~\citep{star2022, mart2023}.  Of course a realistic
cosmological model must include bradyonic (slower-than-light) matter and
radiation (i.e., photonic) constituents.  The former would engender a density
contribution for which the second term on the right-hand sides of
Eqs.~\eqref{eq02a} and~\eqref{eq02b}---the term proportional to~$1/a^3$---is
positive; for the latter that term is zero.  As long as the combined densities
yield a net negative $1/a^3$~term, the model is
tachyon-dominated~\citep{star2022} and evolves as detailed here.

For parameter values for which $A>B/c$ holds, this model undergoes ``inflected''
expansion similar, but not identical, to that of current dark-energy-dominated
or $\Lambda$CDM spacetimes, although this model features no dark energy.
Scale factor~$a(t)$ or~$a(\eta)$ expands from an initial singularity
($a=0$) at a decreasing rate, reaching a minimum expansion rate at
$a=A^2/B$, thence\footnote{This transition is sometimes called the ``cosmic
jerk,'' although that term should propertly refer to the full time dependence
of the expansion acceleration, i.e., the third time derivative~$d^3a/dt^3$
of the scale factor.} expanding at an accelerating rate which asymptoticallly
approaches $da/dt\to c$~\citep{star2022}.  The \textit{deceleration parameter}
for this model is given by
\begin{equation}
\label{eq03}
\begin{aligned}[b]
q&\equiv-\left(\frac{1}{a}\frac{da}{dt}\right)^{-2}\frac{1}{a}
\frac{d^2a}{dt^2}\\
&=1-\frac{a}{(da/d\eta)^2}\,\frac{d^2a}{d\eta^2}\\
q&=\frac{A^2c^2-ABc\sinh\eta+B^2(\cosh\eta-1)}
{(Ac\cosh\eta-B\sinh\eta)^2}\\
\end{aligned}
\end{equation}
This takes value $q=1$ at $\eta=0$, remaining positive (indicating
deceleration) until the model passes through its minimum expansion rate.
It takes negative values (for acceleration) thereafter, approaching zero
for $\eta\to+\infty$ as the expansion rate approaches its constant asymptotic
value~\citep{mart2023}.  This behavior distinguishes the open
tachyon-dominated model from $\Lambda$CDM models, for which the expansion
rate asymptotically accelerates exponentially in comoving time.

The distance-redshift relation for this model follows from spacetime
metric~\eqref{eq01a}--\eqref{eq01c}.  The metric distance~$\ell$ of a
source at radial (angular) coordinate~$\chi$, from which a light signal
is received at the origin at current conformal time~$\eta_0$, is given
by~\citep{mart2023}
\begin{equation}
\label{eq04}
\begin{aligned}[b]
\ell(z)&=a_0\chi\\
&=a_0\,\left[\eta_0-\alpha-\sinh^{-1}\left(
\frac{a_0\,\cosh\alpha}{(A/c)(1+z)}-\sinh\alpha\right)\right]\ ,\\
\end{aligned}
\end{equation}
in terms of redshift~$z$ defined by $1+z\equiv a_0/a(\eta_0-\chi)$,
with parameter $\alpha\equiv\tanh^{-1}[B/(Ac)]$.  The measured {\it luminosity
distance\/}~\citep{mtw1973,lppt1975} is given by
\begin{equation}
\label{eq05}
\begin{aligned}[b]
D_L(z)&\equiv\left(\dfrac{\cal L}{4\pi S}\right)^{1/2}\\
&=(1\times10^{-8}~\hbox{Gpc})\,\exp\left(\dfrac{\ln(10)\,(m-M)}{5}\right)\\
&=a_0\,(1+z)\,\sinh[\ell(z)/a_0]\\
&=a_0\,(1+z)\,\sinh\left[\eta_0-\alpha-\sinh^{-1}\left(
\frac{a_0\,\cosh\alpha}{(A/c)(1+z)}-\sinh\alpha\right)\right]\\
\end{aligned}
\end{equation}
in terms of measured intensity~$S$, source luminosity~${\cal L}$, apparent
magnitude~$m$, and absolute magnitude~$M$.  This relation includes redshift
and time-dilation effects represented by the factor~$(1+z)$, and curvature
corrections included via the $\sinh$ function appropriate to this open model.

All features of the model follow from the values of three parameters, e.g.,
the current curvature radius or scale factor~$a_0$, the current conformal
time~$\eta_0$, and the dimensionless ratio determined by~$\alpha$.  These
values can be extracted by fitting relation~\eqref{eq05} to distance~$D_L$
and redshift~$z$ measurements for suitable collections of Type~Ia supernovae.
 
\section{Supernova Data}\label{sec03}

A number of groups of astronomers have collected magnitude/redshift
(distance/redshift) data on Type~Ia supernovae since Riess {\it et
al\/}~\citep{ries1995}, Perlmutter {\it et al\/}~\citep{perl1998},
and Garnavich {\it et al\/}~\citep{garn1998} established that these objects
could serve as ``standard candles'' for cosmological distance measurements.
These groups include Amanullah {\it et al\/} 2010~\citep{aman2010},
Betoule {\it et al\/} 2014~\citep{beto2014}, Clocchiatti {\it et al\/}
2006~\citep{cloc2006}, Folatelli {\it et al\/} 2010~\citep{fola2010},
Freedman {\it et al\/} 2019~\citep{free2019}, Hamuy {\it et al\/}
2020~\citep{hamu2020}, Jha {\it et al\/} 2006~\citep{jha2006},
Kattner {\it et al\/} 2012~\citep{katt2012}, Kessler {\it et al\/}
2009~\citep{kess2009}, Krisciunas {\it et al\/} 2005~\citep{kris2005},
Rest {\it et al\/} 2014~\citep{rest2014}, Riess {\it et al\/}
1998~\citep{ries1998}, Riess {\it et al\/} 2004~\citep{ries2004},
Riess {\it et al\/} 2005~\citep{ries2005}, Riess {\it et al\/}
2011~\citep{ries2011}, Scolnic {\it et al\/} 2017~\citep{scol2017},
Suzuki {\it et al\/} 2011~\citep{suzu2011}, Tonry {\it et al\/}
2003~\citep{tonr2003}, Walker {\it et al\/} 2015~\citep{walk2015},
and Williams {\it et al\/} 2003~\citep{will2003}.  This is certainly
not an exhaustive list; a complete bibliography of the subject would
require a separate and extensive work.

We have assembled two sets of supernova data from the Riess {\it et al\/}
2004~\citep{ries2004} and Scolnic {\it et al\/} 2017~\citep{scol2017}
compilations; we refer to these henceforth as the Riess data set and the
Scolnic data set. Luminosity distances are obtained from magnitude data
or ``distance moduli'' $\mu\equiv m-M$ as indicated in Eq.~\eqref{eq05}:
\begin{subequations}
\begin{equation}
\label{eq06a}
D_L=(1\times10^{-8}~\hbox{Gpc})\,\exp\left(\dfrac{\ln(10)\,\mu}{5}\right)
\end{equation}
with uncertainties
\begin{equation}
\label{eq06b}
\sigma_D=D_L\,\dfrac{\ln(10)\,\sigma_\mu}{5}\ .
\end{equation}
\end{subequations}
determined by ordinary error propagation.

Data entries from the Riess data set are shown in Table~\ref{tbl01}.
\begin{table}[h]
\begin{center}
\caption{Redshift/Distance Data for Type~Ia Supernovae
from Riess Data Set\label{tbl01}}
\begin{tabular}{cccccc}
\\\tableline\tableline
Name & $z$ & $\mu$ & $D_L$, Gpc & $\sigma_D$, Gpc & Source \\
\tableline
1990T & 0.0400 & 36.38	& 0.1888 & 0.017 & Riess2004 \\
1990af	& 0.050 & 36.84 & 0.2333 & 0.023 & Riess2004 \\
1990O & 0.0307	& 35.90 & 0.1514 & 0.014 & Riess2004 \\
1991S & 0.0560	& 37.31 & 0.2897 & 0.024 & Riess2004 \\
1991U & 0.0331	& 35.54 & 0.1282 & 0.012 & Riess2004 \\
1991ag & 0.0141 & 34.13 & 0.0670 & 0.008 & Riess2004 \\
1992J & 0.0460 & 36.35	& 0.1862 & 0.018 & Riess2004 \\
1992P & 0.0265 & 35.64 & 0.1343 & 0.012 & Riess2004 \\
1992aq & 0.101 & 38.73 & 0.5572 & 0.051 & Riess2004 \\
1992ae & 0.075 & 37.77 & 0.3581 & 0.031 & Riess2004 \\
\tableline
\end{tabular}
\end{center}
\end{table}
The complete table for this data set includes 186 entries.

Data entries from the Scolnic data set are shown in Table~\ref{tbl02}.
\begin{table}[h]
\begin{center}
\caption{Redshift/Distance Data for Type~Ia Supernovae
from Scolnic Data Set\label{tbl02}}
\begin{tabular}{cccccc}
\\\tableline\tableline
Name & $z$ & $\mu$ & $D_L$, Gpc & $\sigma_D$, Gpc & Source \\
\tableline
03D1au & 0.50349 & 42.2688 & 2.8429 & 0.1550 & ScolnicG10 \\
03D1ax & 0.49520 & 42.2277 & 2.7896 & 0.1507 & ScolnicG10 \\
03D1co & 0.67820 & 43.3714 & 4.7237 & 0.4285 & ScolnicG10 \\
03D1ew & 0.86720 & 43.6898 & 5.4697 & 0.4468 & ScolnicG10 \\
03D1fq & 0.79920 & 43.7000 & 5.4954 & 0.4495 & ScolnicG10 \\
03D3ay & 0.37129 & 41.6234 & 2.1119 & 0.1215 & ScolnicG10 \\
03D3bl & 0.35568 & 41.4494 & 1.9493 & 0.1026 & ScolnicG10 \\
03D4ag & 0.28391 & 40.7462 & 1.4101 & 0.0701 & ScolnicG10 \\
03D4au & 0.46691 & 42.5725 & 3.2696 & 0.2132 & ScolnicG10 \\
03D4cx & 0.94791 & 43.7447 & 5.6097 & 0.4232 & ScolnicG10 \\
\tableline
\end{tabular}
\end{center}
\end{table}
The complete table for this data set includes entries for 1048 supernovae.
\vfil\eject
\section{Open Tachyon-Dominated Model Fitted to Supernova Data}\label{sec04}

The open tachyon-dominated model is tested, and the predictions of the model
are obtained, by fitting distance-redshift relation~\eqref{eq05} to the sets
of supernova data.  The statistical measure
\begin{equation}
\label{eq07}
\chi^2\equiv\sum_i^N\dfrac{[D_{Li}^{\rm (obs)}-D_L(z_i)]^2}{\sigma_{Di}^2}\ ,
\end{equation}
with $N$ the number of data points, distance~$D_{Li}^{\rm (obs)}$,
redshift~$z_i$, and uncertainty~$\sigma_{Di}$ from the data set, and
$D_L(z_i)$ from relation~\eqref{eq05}, is minimized by adjusting the
parameters~$a_0$, $\eta_0$, and~$\alpha$.  The best-fit values of these
three parameters are obtained via a Levenberg-Marquardt minimization
program~\citep{pftv1986a}; all other features of the model are calculated
from these.  The values and uncertainties of all the parameters are obtained
via a straightforward Monte Carlo calculation:  Twenty iterations of the
fitting program are carried out, with normally-distributed variations of
standard deviation~$\sigma_{Di}$ introduced into each data
point.~\citep{pftv1986b}  Means and standard deviations for each parameter
value are taken from these sets of results.~\citep{pftv1986c}

\begin{figure*}
\includegraphics{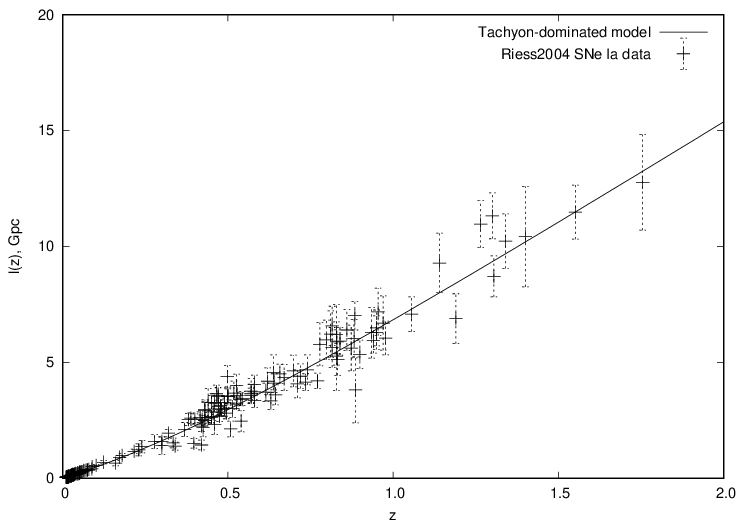}
\caption{\label{f01}Distance-redshift relation for the open tachyon-dominated
model fitted to the Riess data set.  The resulting best-fit parameters are given
by Eqs.~\eqref{eq08a}--\eqref{eq08c}.}
\end{figure*}

\begin{figure*}
\includegraphics{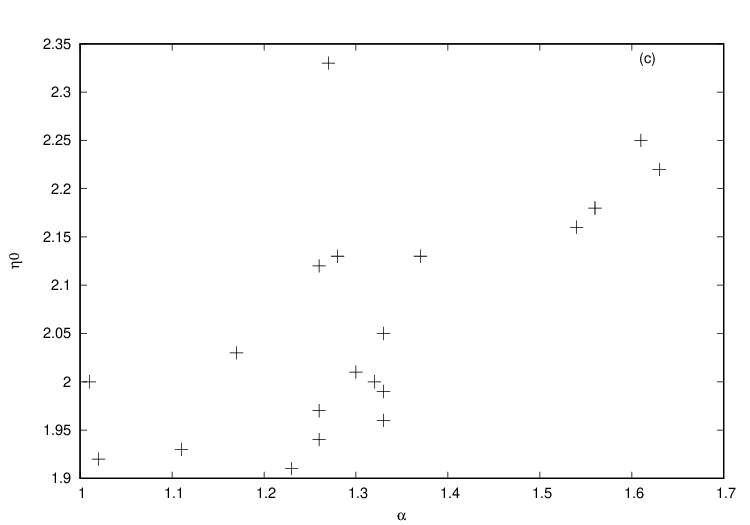}
\caption{\label{f02}Scatter plots showing results of the iterations of the
Levenberg-Marquardt $\chi^2$-minimization program, fitting the Riess data set,
for model parameters (a)~$a_0$ and~$\eta_0$, (b)~$a_0$ and~$\alpha$, and
(c)~$\eta_0$ and~$\alpha$.}
\end{figure*}

The fit to the Riess data set ($N=186$) is shown in Figure~\ref{f01}.
Results of the Monte Carlo calculations for model parameters~$a_0$, $\eta_0$,
and~$\alpha$ are illustrated in Fig.~\ref{f02}.
   
\begin{figure*}
\includegraphics{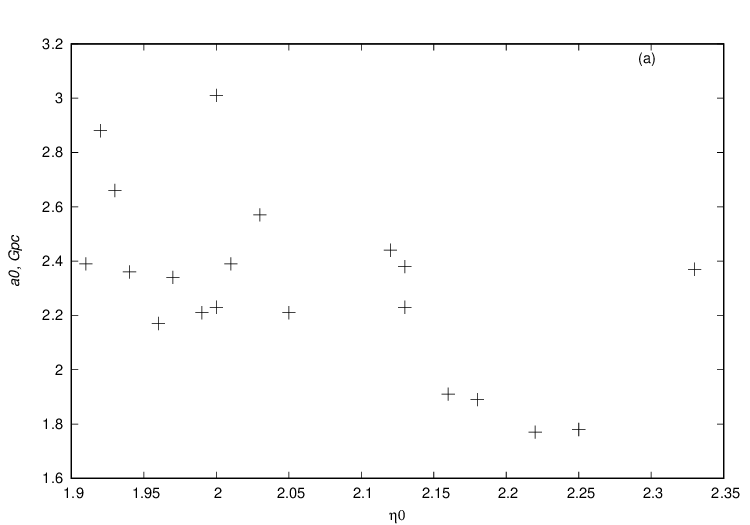}
\end{figure*}
\begin{figure*}
\includegraphics{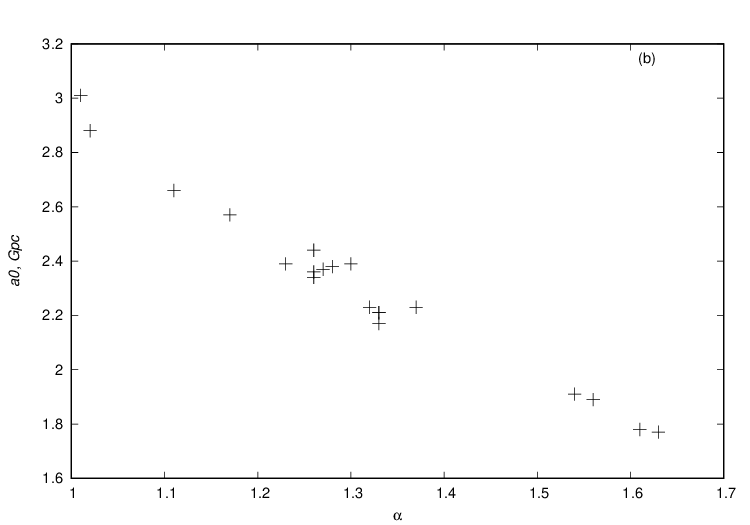}
\end{figure*}

The best-fit model parameters from this calculation are
\begin{subequations}
\begin{equation}
\label{eq08a}
a_0=(2.31\pm0.33)~\hbox{Gpc}\ ,
\end{equation}
\begin{equation}
\label{eq08b}
\eta_0=2.06\pm0.12\ ,
\end{equation}
and
\begin{equation}
\label{eq08c}
\alpha=1.31\pm0.17\ .
\end{equation}
Model features following from these values include present-time Hubble parameter
\begin{equation}
\label{eq08d}
\begin{aligned}[b]
H_0&=\dfrac{1}{a_0}\left(\dfrac{da}{dt}\right)_0\\
&=\dfrac{c}{a_0}\,\dfrac{\cosh(\eta_0-\alpha)}
{\sinh(\eta_0-\alpha)+\sinh\alpha}\\
H_0&=(66.6\pm1.5)~\hbox{km/s/Mpc}\ ,\\
\end{aligned}
\end{equation}
current age of the Universe
\begin{equation}
\label{eq08e}
\begin{aligned}[b]
t_0&=\dfrac{a_0}{c}\,\dfrac{\cosh(\eta_0-\alpha)-\cosh\alpha+\eta_0\,\sinh\alpha}
{\sinh(\eta_0-\alpha)+\sinh\alpha}\\
&=(8.35\pm0.67)~\hbox{Gyr}\\
\end{aligned}
\end{equation}
in comoving time, and cosmic-jerk (inflection point) redshift
\begin{equation}
\label{eq08f}
\begin{aligned}[b]
z_j&=\dfrac{\sinh\alpha\,\sinh(\eta_0-\alpha)-1}{\cosh^2\alpha}\\
&=0.102\pm0.058\ .
\end{aligned}
\end{equation}
\end{subequations}
The best-fit value of~$\chi^2$ for this set of iterations is
$\chi^2=471\pm66$. or $2.57\pm0.36$ per degree of freedom.
\vfil\eject

\begin{figure*}
\includegraphics{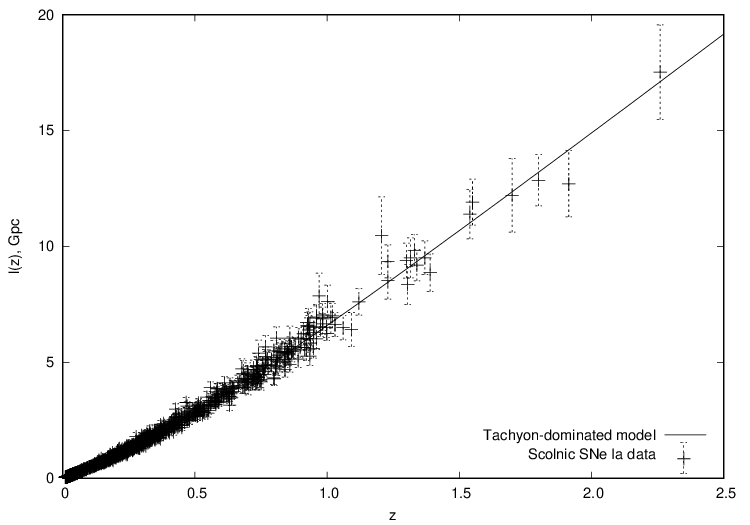}
\caption{\label{f03}Distance-redshift relation for the open tachyon-dominated
model fitted to the Scolnic data set.  The resulting best-fit parameters are
given by Eqs.~\eqref{eq09a}--\eqref{eq09c}.}
\end{figure*} 
\begin{figure*}
\includegraphics{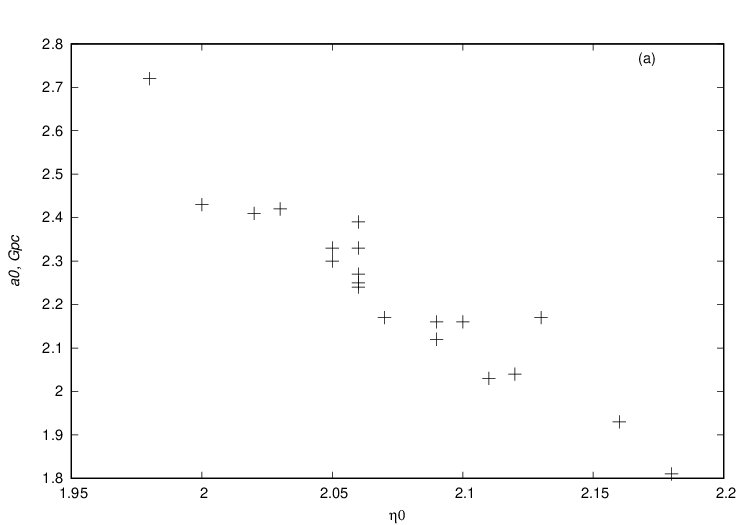}
\end{figure*}
\begin{figure*}
\includegraphics{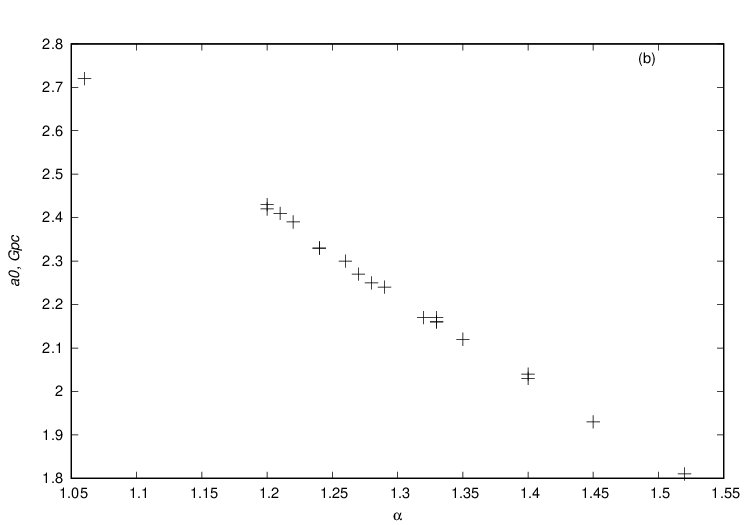}
\end{figure*}
\vfil\eject

\begin{figure*}
\includegraphics{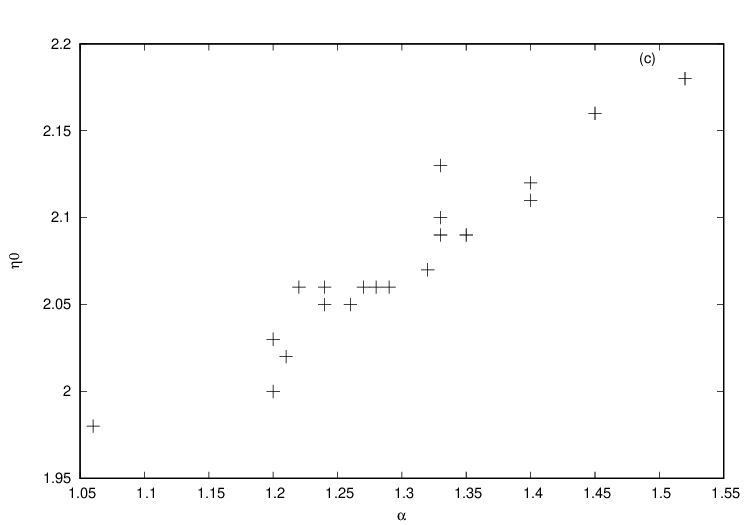}
\caption{\label{f04}Scatter plots showing results of the iterations of the
Levenberg-Marquardt $\chi^2$-minimization program, fitting the Scolnic data set,
for model parameters (a)~$a_0$ and~$\eta_0$, (b)~$a_0$ and~$\alpha$, and
(c)~$\eta_0$ and~$\alpha$.}
\end{figure*}

\vfil\eject

The more ambitious fit to the Scolnic data set ($N=1048$) is shown in
Figure~\ref{f03}.  Results of this Monte Carlo calculation are illustrated
in Fig.~\ref{f04}.  The best-fit parameters from this calculation are
\begin{subequations}
\begin{equation}
\label{eq09a}
a_0=(2.23\pm0.20)~\hbox{Gpc}\ ,
\end{equation}
\begin{equation}
\label{eq09b}
\eta_0=2.07\pm0.05\ ,
\end{equation}
and
\begin{equation}
\label{eq09c}
\alpha=1.30\pm0.10\ .
\end{equation}
These imply the corresponding model features
\begin{equation}
\label{eq09d}
H_0=(69.6\pm0.4)~\hbox{km/s/Mpc}\ ,
\end{equation}
\begin{equation}
\label{eq09e}
t_0=(8.15\pm0.36)~\hbox{Gpc}
\end{equation}
and
\begin{equation}
\label{eq09f}
z_j=0.115\pm0.011\ .
\end{equation}
\end{subequations}
The best-fit value of~$\chi^2$ for these iterations is $\chi^2=2080\pm80$.
or $1.99\pm0.08$ per degree of freedom.

Features of the tachyon gas can be inferred from these parameter values.
If that gas is taken to constitute the entire cosmological fluid, then
the current invariant-mass density from Eqs.~\eqref{eq02a} and~\eqref{eq02b}
is
\begin{equation}
\label{eq10}
\begin{aligned}[b]
{\cal M}_0c^2&=\dfrac{3c^2}{8\pi G}\,\dfrac{2B}{a_0^3}\\
&\doteq 4.07\times 10^{-9}~\hbox{J/m}^3\ ,\\
\end{aligned}
\end{equation}
using values from the Scolnic-data fit~\eqref{eq09a}--\eqref{eq09c}.  The current
total energy density is
\begin{equation}
\label{eq11}
\begin{aligned}[b]
\rho_0&=\dfrac{3c^2}{8\pi G}\,\dfrac{A^2}{a_0^4}-{\cal M}_0c^2\\
&\doteq -2.24\times 10^{-9}~\hbox{J/m}^3\\
\end{aligned}
\end{equation}
for the same parameter values.  The negative value is as expected, since
the current scale factor~$a_0$ is beyond the inflection-point value
$a_j\doteq2.00~\hbox{Gpc}$ and the smaller value~$a_j/2$, at which the
density crosses from positive to negative values due to expansion against
the high pressure of the tachyon gas.~\citep{star2022}  In a more realistic
model the tachyonic contribution to the $A$ term in the density would be less,
since bradyonic and photonic constituents also contribute to that term.
The tachyonic contribution to the $B$~term would be greater in magnitude,
since a bradyonic constituent would contribute with the opposite sign.
Hence invariant-mass density~\eqref{eq10} is a lower bound, and
density~\eqref{eq11} an upper bound, to these properties of the tachyon
gas dominating the model.

\section{Conclusions}\label{sec05}

These results indicate that the open tachyon-dominated model might be a viable
alternative to the currently-standard $\Lambda$CDM models dominated by some
form of dark energy.  The best-fit parameters obtained from the two data
sets, Eqs.~\eqref{eq08a}--\eqref{eq08f} and Eqs.~\eqref{eq09a}--\eqref{eq09f},
are comfortably consistent with one another.  The minimal $\chi^2$~values
are only slightly larger than ideal.  (This might be due e.g., to a slight
underestimate of distance uncertainties.)  Features of the model such
as~$H-0$ and~$t_0$ values are similar, but not identical, to those obtained
from other models.  And the tachyonic model engenders the fits illustrated
in Figs.~\ref{f01} and~\ref{f02} with fewer parameters---here, three---than
a $\Lambda$CDM model, which must include at least one further parameter for
the dark-energy density or cosmological constant.

The next stages in the exploration of this model are to compare the consistency
of the parameters obtained by fitting its predictions to other cosmological
observations.  These might include the microwave background,~\citep{gopa2024}
quasar microlensing,~\citep{john2024} acoustic waves,~\citep{suth2024} the
Lyman~$\alpha$ forest,~\citep{lash2024} and other phenomena.  Such works are
in progress at this time.

\clearpage


\begin{thebibliography}{}
\bibitem[Starke \& Redmount(2022)]{star2022} Starke, J., \& Redmount, I. H.
   2022, arXiv:1905.13557\_v2, International Journal of Modern Physics A,
   37, 2250162
\bibitem[Martin \& Redmount(2023)]{mart2023} Martin, A. C., \& Redmount,
   I. H. 2023, arXiv:1904.07316, International Journal of Modern Physics D, 32,
   2350028
\bibitem[Misner, Thorne, \& Wheeler(1973)]{mtw1973} Misner, C. W.,
   Thorne, K. S., \& Wheeler, J. A., Gravitation, San Francisco, CA:
   Freeman, 782--785
\bibitem[Lightman, Press, Price, \& Teukolsky(1975)]{lppt1975} Lightman,
   A. P., Press, W. H., Price, R. H., \& Teukolsky, S. A., Problem Book in
   Relativity and Gravitation, Princeton, NJ: Princeton, 114 and 527--528
\bibitem[Riess et al.(1995)]{ries1995} Riess, A. G., Press, W. H., \& Kirshner,
   R. P. 1995, \apj, 438, L17
\bibitem[Perlmutter et al.(1998)]{perl1998}Perlmutter, S., Aldering, G.,
   Della Valle, M., Deustua, S., Ellis, R. S., Fabbro, S., Fruchter, A.,
   Goldhaber, G., Groom, D. E., Hook, I. M., Kim, A. G., Kim, M. Y.,
   Knop, R. A., Lidman, C., McMahon, R. G., Nugent, P., Pain, R., Panagia, N.,
   Pennypacker, C. R., Ruiz-Lapuente, P., Schaefer, B., \& Walton, N. 1998,
   Nature, 391, 51
\bibitem[Garnavich et al.(1998)]{garn1998} Garnavich, P. M.,
   Kirshner, R. P., Challis, P., Tonry, J., Gilliland, R. L.,
   Smith, R. C., Clocchiatti, A., Diercks, A., Filippenko, A. V.,
   Hamuy, M., Hogan, C. J., Beibundgut, B., Phillips, M. M.,
   Reiss, D., Riess, A. G., Schmidt, B. P., Schommer, R. A.,
   Spyromilio, J., Stubbs, C., Suntzeff, N. B., \& Wells, L. 1998,
   \apjl, 493, L53
\bibitem[Amanullah et al.(2010)]{aman2010} Amanullah, R., Lidman, C., Rubin,
   D., Aldering, G., Astier, P., Barbary, K., Burns, M. S., Conley, A.,
   Dawson, K. S., Deustua, S. E., Doi, M., Fabbro, S., Faccioli, L., Fakhouri,
   H. K., Folatelli, G., Fruchter, A. S., Furusawa, H., Garavini, G.,
   Goldhaber, G., Goobar, A., Groom, D. E., Hook, I., Howell, D. A.,
   Kashikawa, N., Kim, A. G., Knop, R. A., Kowalski, M., Linder, E., Meyers,
   J., Morokuma, T., Nobili, S., Nordin, J., Nugent, P. E., \"{O}stman, L.,
   Pain, R., Panagia, N., Perlmutter, S., Raux, J., Ruiz-Lapuente, P.,
   Spadafora, A. L., Strovink, M., Suzuki, N., Wang, L., Wood-Vasey, W. M.,
   \& Yasuda, N. 2010, \apj, 716, 712--738
\bibitem[Betoule et al.(2014)]{beto2014} Betoule, M., Kessler, R., Guy, J.,
   Mosher, J., Hardin, D., Biswas, R., Astier, P., El-Hage, P., Konig, M.,
   Kuhlmann, S., Marriner, J., Pain, R., Regnault, N., Balland, C.,
   Bassett, B. A., Brown, P. J., Campbell, H., Carlberg, R. G.,
   Cellier-Holzem, F., Cinabro, D., Conley, A., D'Andrea, C. B., DePoy, D. L.,
   Doi, M., Ellis, R. S., Fabbro, S., Filippenko, A. V., Foley, R. J.,
   Frieman, J. A., Fouchez, D., Galbany, L., Goobar, A., Gupta, R. R.,
   Hill, G. J., Hlozek, R., Hogan, C. J., Hook, I. M., Howell, D. A.,
   Jha, S. W., LeGuillou, L., Leloudas, G., Lidman, C., Marshall, J. L.,
   M\"{o}ller, A., Mour\~{a}o, A. M., Neveu, J., Nichol, R., Olmstead, M. D.,
   Palanque-Delabrouille, N., Perlmutter, S., Prieto, J. L., Pritchet, C. J.,
   Richmond, M., Riess, A. G., Ruhlmann-Kleider, V., Sako, M., Schahmaneche,
   K., Schneider, D. P., Smith, M., Sollerman, J., Sullivan, M., Walton, N. A.
   \& Wheeler, C. J.2014, Astronomy \& Astrophysics, 568, A22
\bibitem[Clocchiatti et al.(2006)]{cloc2006} Clocchiatti, A., Schmidt, B. P.,
   Filippenko, A. V., Challis, P., Coil, A. L., Covarrubias, R., Diercks, A.,
   Garnavich, P., Germany, L., Gilliland, R., Hogan, C., Jha, S., Kirshner,
   R. P., Leibundgut, B., Leonard, D., Li, W., Matheson, T., Phillips, M. M.,
   Prieto, J. L., Reiss, D., Riess, A. G., Schommer, R., Smith, R. C.,
   Soderberg, A., Spyromilio, J., Stubbs, C., Suntzeff, N. B., Tonry, J. L.,
   \& Woudt, P. 2005, arXiv:astro-ph/0510155v1, \apj, 642, 1
\bibitem[Folatelli et al.(2010)]{fola2010} Folatelli, G., Phillips, M. M.,
   Burns, C. R., Contreras, C., Hamuy, M., Freedman, W. L., Persson, S. E.,
   Stritzinger, M., Suntzeff, N. B., Krisciunas, K., Boldt, L., Gonz\'{a}lez,
   S., Krzeminski, W., Morrell, N., Roth, M., Salgado, F., Madore, B. F.,
   Murphy, D., Wyatt, P., Li, W., Filippenko, A. V., \& Miller, N. 2010,
   arXiv:0910.3317v1, The Astronomical Journal, 139, 120
\bibitem[Freedman et al.(2019)]{free2019} Freedman, W. L., Madore, B. F., Hatt,
   D., Hoyt, T. J., Jang, I. S., Beaton, R. L., Burns, C. R., Lee, M. G.,
   Monson, A. J., Neeley, J. R., Phillips, M. M., Rich, J. A., \& Seibert, M.
   2019, arXiv:1907.05922v1, \apj, 882, 34
\bibitem[Hamuy et al.(2020)]{hamu2020} Hamuy, M., Cartier, R., Contreras, C.,
   \& Suntzeff, N. B. 2020, arXiv:2009.10279v2, \mnras, 500, 1095
\bibitem[Jha et al.(2006)]{jha2006} Jha, S., Riess, A. G., \& Kirshner, R. P.
   2006, arXiv:astro-ph/0612666v1, \apj, 659, 122
\bibitem[Kattner et al.(2012)]{katt2012} Kattner, S., Leonard, D. C., Burns,
   C. R., Phillips, M. M., Folatelli, G., Morrell, N., Stritzinger, M. D.,
   Hamuy, M., Freedman, W. L., Persson, S. E., Roth, M., \& Suntzeff, N. B.
   2012, arXiv:1201.2913v2, \pasp, 124, 114
\bibitem[Kessler et al.(2009)]{kess2009} Kessler, R., Becker, A. C., Cinabro,
   D., Vanderplas, J., Frieman, J. A., Marriner, J., Davis, T. M., Dilday, B.,
   Holtzman, J., Jha, S. W., Lampeitl, H., Sako, M., Smith, M., Zheng, C.,
   Nichol, R. C., Bassett, B., Bender, R., Depoy, D. L., Doi, M., Elson, E.,
   Filippenko, A. V., Foley, R. J., Garnavich, P. M., Hopp, U., Ihara, Y.,
   Ketzeback, W., Kollatschny, W., Konishi, K., Marshall, J. L., McMillan,
   R. J., Miknaitis, G., Morokuma, T., M\"{o}rtsell, E., Pan, K., Prieto,
   J. L., Richmond, M. W., Riess, A. G., Romani, R., Schneider, D. P.,
   Sollerman, J., Takanashi, N., Tokita, K., van der Heyden, K., Wheeler,
   J. C., Yasuda, N., \& York, D. 2009, arXiv:0908.4274v1, \apjs, 185, 32
\bibitem[Krisciunas et al.(2005)]{kris2005} Krisciunas, K., Garnavich, P. M.,
   Challis, P., Prieto, J. L., Riess, A. G., Barris, B., Aguilera, C., Becker,
   A. C., Blondin, S., Chornock, R., Clocchiatti, A., Covarrubias, R.,
   Filippenko, A. V., Foley, R. J., Hicken, M., Jha, S., Kirshner, R. P.,
   Leibundgut, B., Li, W., Matheson, T., Miceli, A., Miknaitis, G., Rest, A.,
   Salvo, M. E., Schmidt, B. P., Smith, R. C., Sollerman, J., Spyromilio,
   J., Stubbs, C. W., Suntzeff, N. B., Tonry, J. L., \& Wood-Vasey, W. M. 2005,
   arXiv:astro-ph/0508681v1, The Astronomical Journal, 130, 2453
\bibitem[Rest et al.(2014)]{rest2014} Rest, A., Scolnic, D., Foley, R. J.,
   Huber, M. E., Chornock, R., Narayan, G., Tonry, J. L., Berger, E., Soderberg,
   A. M., Stubbs, C. W., Riess, A., Kirshner, R. P., Smartt, S. J., Schlafly, E.,
   Rodney, S., Botticella, M. T., Brout, D., Challis, P., Czekala, I., Drout, M.,
   Hudson, M. J., Kotqk, R., Leibler, c., Lunnan, R., Marian, G. H., McCrum, M.,
   Milisavljevich, D., Pastorello, A., Sanders, N. E., Smith, K., Stafford, E.,
   Thilker, D., Valenti, S., Wood-Vasey, W. M., Zheng, Z., Burgett, W. S.,
   Chambers, K. C., Denneau, L., Draper, P. W., Flewelling, H., Hodapp, K. W.,
   Kaiser, N., Kudritzki, R.-P., Magnier, E. A., Metcalfe, N., Price, P. A.,
   Sweeney, W., Wainscoat, R., \& Waters, C. 2014, \apj, 795 , 44
\bibitem[Riess et al.(1998)]{ries1998} Riess, A. G., Filippenko, A. V., Challis,
   P., Clocchiatti, A., Diercks, A., Garnavich, P. M., Gilliland, R. L., Hogan,
   C. J., Jha, S., Kirshner, R. P., Leibundgut, B., Phillips, M. M., Reiss, D.,
   Schmidt, B. P., Schommmer, R. A., Smith, R. C., Spyromilio, J., Stubbs, C.,
   Suntzeff, N. B., \& Tonry, J. 1998, arXiv:astro-ph/9805201v1, The Astronomical
   Journal, 116, 1009
\bibitem[Riess et al.(2004)]{ries2004} Riess, A. G., Strolger, L.-G., Tonry, J.,
   Casertano, S., Ferguson, H. C., Mobasher, B., Challis, P., Filippenko, A. V.,
   Jha, S., Li, W., Chornock, R., Kirshner, R. P., Leibundgut, B., Dickinson, M.,
   Livio, M., Giavalisco, M., Steidel, C. C., Benitez, N., \& Tavetanov, Z.
   2004, arXiv:astro-ph/0402512v2, \apj, 607, 665
\bibitem[Riess et al.(2005)]{ries2005} Riess, A. G., Li, W., Stetson, P. B.,
   Filippenko, A. V., Jha, S., Kirshner, R. P., Challis, P. M., Garnavich, P. M.,
   \& Chornock, R. 2005, arXiv:astro-ph/0503159v2, \apj, 627, 579
\bibitem[Riess et al.(2011)]{ries2011} Riess, A. G., Macri, L., Casertano, S.,
   Lampeitl, H., Ferguson, H. C., Filippenko, A. V., Jha, S. W., Li, W., \&
   Chornock, R. 2011, arXiv:1103.2976v1, \apj, 730, 119
\bibitem[Scolnic et al.(2017)]{scol2017} Scolnic, D. M., Jones, D. O., Rest, A.,
   Pan, Y. C., Chornock, R., Foley, R. J., Huber, M. E., Kessler, R., Narayan, G.,
   Riess, A. G., Rodney, S., Berger, E., Brout, D. J., Challis, P. J., Drout, M.,
   Finkbeiner, D., Lunnan, R., Kirshner, R. P., Sanders, N. E., Schlafly, E.,
   Smartt, S., Stubbs, C. W., Tonry, J., Wood-Vasey, W. M., Foley, M., Hand, J.,
   Johnson, E., Bergett, W. S., Chambers, K. C., Draper, P. W., Hodapp, K. W.,
   Kaiser, N., Kudritzki, R. P., Magnier, E. A., Metcalfe, N., Bresolin, F.,
   Gall, E., Kotak, R., McCrum, M., \& Smith, K. W. 2018, arXiv:1710.00845v3,
   \apj, 859, 101
\bibitem[Suzuki et al.(2011)]{suzu2011} Suzuki, N., Rubin, D., Lidman, C.,
   Aldering, G., Amanullah, R., Barbary, K., Barrientos, L. F., Botyanszki, J.,
   Drodwin, M., Connolly, N., Dawson, K. S., Dey, A., Doi, M., Donahue, M.,
   Deustua, S., Eisenhardt, P., Ellingson, E., Faccioli, L., Fadeyev, V.,
   Fakhouri, H. K., Fruchter, A. S., Gilbank, D. G., Gladders, M. D., Goldhaber, G.,
   Gonzalez, A. H., Goobar, A., Gude, A., Hattori, T., Hoekstra, H., Hsiao, E.,
   Huang, X., Ihara, Y., Jee, M. J., Johnston, D., Kashikawa, N., Koester, B.,
   Konishi, K., Kowalski, M., Linder, E. V., Lubin, L., Melbourne, J., Meyers, J., 
   Morokuma, T., Munshi, F., Mullis, C., Oda, T., Panagia, N., Perlmutter, S.,
   Postman, M., Pritchard, T., Rhodes, J., Ripoche, P., Takanashi, N., Tokita, K.,
   Wagner, M., Wang, L., Yasuda, N., \& Yee, H. K. C. 2011, arXiv:1105.3470v1,
   \apj, 746, 85
\bibitem[Tonry et al.(2003)]{tonr2003} Tonry, J. L., Schmidt, B. P., Barris, B.,
   Candia, P., Challis, P., Clocchiatti, A., Coil, A. L., Filippenko, A. V.,
   Garnavich, P., Hogan, C., Holland, S. T., Jha, S., Kirshner, R. P., Krisciunas, K.,
   Leibundgut, B., Li, W., Matheson, T., Phillips, M. M., Riess, A. G., Schommer, R.,
   Smith, R. C., Sollerman, J., Spyromilio, J., Stubbs, C. W., \& Suntzeff, N. B.
   2003, arXiv:astro-ph/0305008v1, \apj, 594, 1
\bibitem[Walker et al.(2015)]{walk2015} Walker, E. S., Baltay, C., Campillay, A.,
   Citrenbaum, C., Contreras, C., Ellman, N., Feindt, U., Gonz\'{a}lez, C.,
   Graham, M. L., Hadjiyska, E., Hsiao, E. Y., Krisciunas, K., McKinnon, R., Ment, K.,
   Norrell, N., Nugent, P., Phillips, M. M., Rabinowitz, D., Rostami, S., Ser\'{o}n,
   J., Stritzinger, M., Sullivan, M., \& Tucker, B. E. 2015, \apjs, 219, 13
\bibitem[Williams et al.(2003)]{will2003} Williams, B. F., Hogan, C. J., Barris, B.,
   Candia, P., Challis, P., Clocchiatti, A., Coil, A. L., Filippenko, A. V., Garnavich,
   P., Kirshner, R. P., Holland, S. T., Jha, S., Krisciunas, K., Leibundgut, B., Li, W.,
   Matheson, T., Maza, J., Phillips, M. M., Riess, A. G., Schmidt, B. P., Schommer,
   R. A., Smith, R. C., Sollerman, J., Spyromilio, J., Stubbs, C., Suntzeff, N. B.,
   \& Tonry, J. L. 2003, arXiv:astro-ph/0310432v1, The Astronomical Journal, 126, 2608   
\bibitem[Press, Flannery, Teukolsky, \& Vetterling(1986a)]{pftv1986a} Press,
   W. H., Flannery, B. P., Teukolsky, S. A., \& Vetterling, W. T., Numerical
   Recipes:  The Art of Scientific Computing, Cambridge, UK: Cambridge,
   523--528
\bibitem[Press et al.(1986b)]{pftv1986b} Press, W. H., et al, op.~cit.,
   202--203
\bibitem[Press et al.(1986c)]{pftv1986c} Press, W. H., et al, op.~cit.,
   529--532
\bibitem[Gopal \& Redmount(2024)]{gopa2024} Gopal, N. \& Redmount, I. H.
   2024, in preparation
\bibitem[Johnson \& Redmount(2024)]{john2024} Johnson, K. \& Redmount, I. H.
   2024, in preparation
\bibitem[Suthan \& Redmount(2024)]{suth2024} Suthan, H. \& Redmount, I. H.
   2024, in preparation
\bibitem[Lashbrook \& Redmount(2024)]{lash2024} Lashbrook, A. \& Redmount,
I. H. 2024, in preparation
\end{thebibliography}
\end{document}